\documentclass[twocolumn,showpacs,preprintnumbers,amsmath,amssymb,prl]{revtex4-1}

\usepackage{graphicx}
\usepackage{dcolumn}
\usepackage{bm}
\usepackage{hyperref}
\usepackage{color}
\usepackage[OT1]{fontenc}

\begin{document}

\title{Avalanche Instability as Nonequilibrium Quantum Criticality}

\author{Xi Chen}
\affiliation{Department of Physics, State University of New York at Buffalo, Buffalo, New York 14260, USA}
\author{Jong E. Han}
\email{jonghan@buffalo.edu}
\affiliation{Department of Physics, State University of New York at Buffalo, Buffalo, New York 14260, USA}

\date{\today}

\begin{abstract} 

A fundamental instability in the nonequilibrium conduction band under a
electric field bias is proposed via the spontaneous emission of coherent
phonons. Analytic theory, supported by numerical calculations,
establishes that the quantum avalanche, an abrupt nonequilibrium
occupation of excited bands, results from the competition between the
collapse of the band minimum via the phonon emission and the dephasing
of the electron with the environment.  The continuous avalanche
transition is a quantum phase transition with the nonequilibrium phase
diagram determined by the avalanche parameter $\beta$, with peculiar
reentrant avalanche domes close to the phase boundary.  We further
confirm the nature of the quantum avalanche with the temperature
dependence. 

\end{abstract}

\maketitle

In the past half-century, materials under strong
electromagnetic field have been extensively studied. In particular, the
resistive phase transition driven by a high electric field has generated
strong research efforts~\cite{bardeenPT,bardeen1989,Ridley,thornehistory,janod}. However, despite
the scientific and technological importance of the phenomena, conceptual
advancement has been limited since it requires an understanding of
many-body dynamics far-from-equilibrium~\cite{aoki}.

The challenge partly comes from the lack of theoretical milestones.
Despite mounting experimental reports~\cite{ong1979,grunerRMP,
janod,zimmers,zhang_nano}, theories have not provided decisive new
insights into outstanding issues. One such problem is resistive
switching, in which the mechanism of the insulator-to-metal transition
by a DC electric field has been debated, as to the electronic or thermal
origin, for many decades without much consensus. Part of the problem is
that theoretical efforts have been often too complex to systematically
relate to well-established equilibrium counterparts. The goal of our
analytic theory is to identify a mechanism of nonequilibrium quantum
transition akin to critical phenomena and provide a conceptual and
transparent framework that could initiate future discussion.

In the past decades, we have asked how electrons overcome the energy gap to
induce dielectric breakdown, mainly within the framework of
Landau-Zener tunneling~\cite{bardeenPT,zener,ong1979}. Despite strong efforts, theories failed to address
the energy-scale discrepancy where the experimental switching fields are
orders of magnitude smaller than theoretical
predictions~\cite{oka2003,sugimoto,eckstein2010,han2018}. In a recent
work~\cite{han_avalanche}, an alternative answer was proposed. 
In materials, electrical resistivity is not
infinite, and, with bias, there exist many charge-carriers present in the
bulk limit, despite with very dilute concentration. 
Once electrons are coupled to an inelastic medium, instability
develops with a uniform electric field, in principle at infinitesimally
small strength, leading to an eventual resistive breakdown of the system at
experimental scales~\cite{han_avalanche}. We show here that, through an
analytic study, the avalanche instability is a subset of the fundamental
instability of a band under bias by analyzing a minimal nonequilibrium
steady-state model.

We introduce a model of a 1-dimensional electron gas coupled to
optical-phonons with electrons subject to a static and uniform electric
field $E$ with the Hamiltonian
\begin{eqnarray}
H(t) & =  & \int
\left[\psi^\dagger(x)\left(\frac{1}{2m}(-i\hbar\partial_x+eEt)^2+\Delta\right)\psi(x)
\right.
\\
& + &
\left.\frac12 \left(p_\varphi(x)^2+\omega_0^2\varphi(x)^2\right)
+ 
g_{\rm ep}\varphi(x)\psi^\dagger(x)\psi(x)
\right]dx
\nonumber
\end{eqnarray}
with the (spinless) electron creation/annihilation operator
$\psi^\dagger(x)$/$\psi(x)$, the Einstein phonon field $\varphi(x)$ of
frequency $\omega_0$ with its
conjugate momentum $p_\varphi(x)$, and the electron-phonon coupling
constant $g_{\rm ep}$. The conduction band is placed $\Delta$ above
the Fermi energy of the particle reservoir. A uniform DC electric field is included as a
vector potential $-eEt\hat{\bf x}$ in the temporal
gauge~\cite{khan_wilkins,kemper}. We use the unit system that
$\hbar=e=k_B=1$, with the Boltzmann constant $k_B$.

The system is coupled to the environment
with electrons and phonons connected to the fermionic and bosonic
baths~\cite{han_avalanche,moire2023,mazzocchi}, respectively.
The importance of phonon baths has been highlighted~\cite{han_avalanche,zhang_chern}, recently.
The fermion reservoirs account for exchange of electrons from bands
outside model (such as substrate) and provides dissipation. More
importantly this mechanism sets the electron lifetime via dephasing from
external sources other than phonons. The hybridization
to the fermion bath is given as $\Gamma$, which we assume to be
independent of energy and to be structureless with infinite bandwidth,
for simplicity. The electrons can deposit their excess energy into phonons
with the scattering rate controlled by the coupling constant $g_{\rm
ep}$, with the excited phonons eventually decaying into an Ohmic
bath~\cite{weiss,khurgin2007}.
 
\begin{figure}
\begin{center}
\rotatebox{0}{\resizebox{3.5in}{!}{\includegraphics{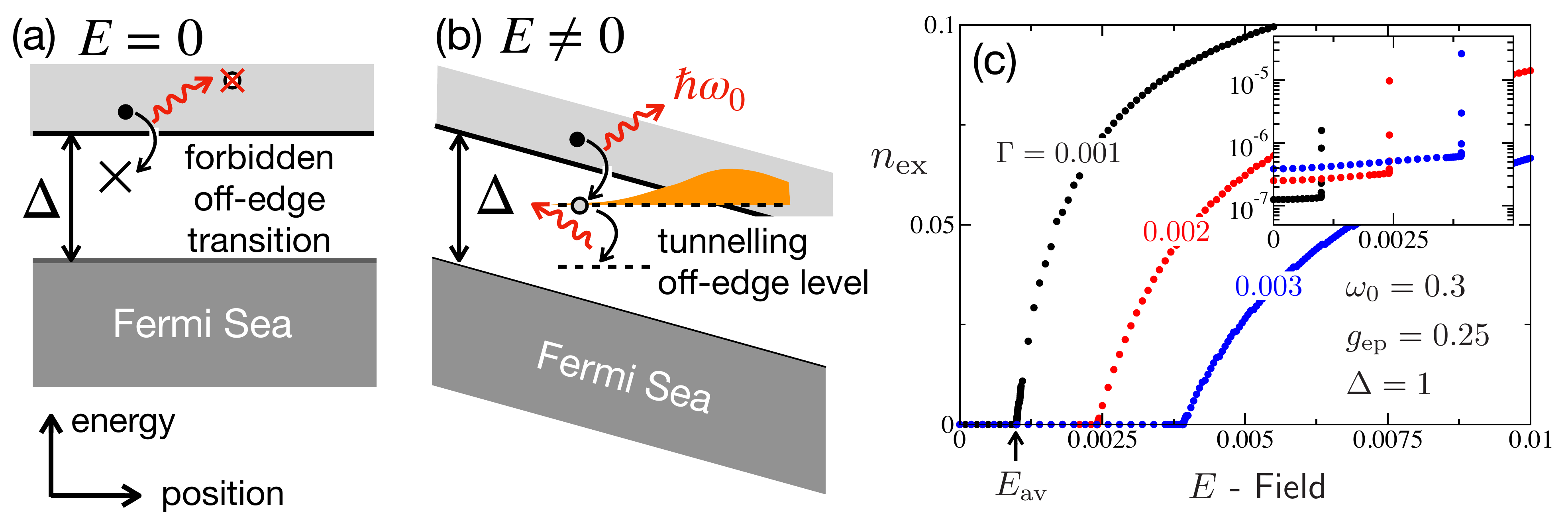}}}
\caption{(a) Energy scheme of a conduction band above the Fermi level by
$\Delta$ at equilibrium. Spontaneous emission of phonon into an
electronic level below the band edge is not allowed. (b) With the electric field $E>0$, the potential slope provides
energy levels tunneling below the band edge, enabling spontaneous transition
into the forbidden region by emitting local phonons.
As the electronic replica state, with its energy lowered by $\omega_0$,
is reinforced by the multiple-phonon processes, an abrupt quantum
transition occurs in an avalanche. (c) Numerical results showing
occupation number $n_{\rm ex}$ of the conduction band as a function of
$E$. The avalanche field $E_{\rm av}$ is an increasing
function of the dephasing rate $\Gamma$, suggesting that the dephasing
competes with the avalanching mechanism. Counter-intuitively, smaller
pre-avalanche occupations led to earlier avalanches, as shown in the
inset.
}
\label{fig1}
\end{center}
\end{figure}

This minimal model has been shown to induce a quantum
avalanche~\cite{han_avalanche} where a phase transition to a strong
nonequilibrium occupation of the band occurs at a small electric-field scale.
The mechanism for the quantum avalanche is as follows. As depicted in
FIG.~\ref{fig1}(a), spontaneous phonon emission does not occur
in the $E=0$ limit due to the absence of states below the band minimum. Even with the
faint line-broadening into the gap due to $\Gamma$, these states
are quite insignificant. However, with a non-zero electric field [see
FIG.~\ref{fig1}(b)], the potential slope provides electronic levels at
any energy. (Here, we temporarily switch to the static gauge with
potential $V(x)=-eEx$ for the sake of argument.) While the off-edge
states are due to the evanescent tail centered at different position as
depicted as the orange envelope function in (b), it allows much
enhanced spontaneous phonon emission compared to (a). This smear of
bandedge due to a uniform electric field is the Franz-Keldysh
effect~\cite{franzkeldysh}. With the replica state generated by a
phonon-emission reinforces the evanescent off-edge state so that it can act
as the reference state that generates the second replica state. The
formation of the multiple replicas requires the phase coherence between the
electron and the phonon throughout, which is limited by the electron
dephasing time. This sets the threshold for the quantum avalanche. We note
that the nature of the transition is spontaneous electronic transition
below the band, instead of the sequential dissipation of excess
electronic energy into phonon quanta.

We emphasize that the following theoretical analysis is confirmed by
fully numerical calculations. As published in several nonequilibrium
dynamical mean-field theory works~\cite{han_avalanche,liprl2015,ligraphene,moire2023},
the dissipation mechanisms are rigorously implemented in self-consistent
calculations. It is essential to include the dissipation on an equal
footing as the system Hamiltonian to ensure numerical convergence. We
detail the numerical procedures for 
the lattice model with the static gauge and with full dissipation
in the Supplementary Materials (SM) for completeness. The tight-binding
parameter $t$ was set to $ta^2=\hbar^2/(2m)=1$ with lattice constant $a$
which is scaled to 1 in the calculation.

Before we present the analytic theory, we discuss numerical evidence for
the quantum avalanche, in FIG.~\ref{fig1}(c). The occupation number of
the band $n_{\rm ex}$ shows a continuous phase transition at the finite electric
field at $E_{\rm av}$. There are two key observations: (1) The
avalanche field $E_{\rm av}$ is almost linearly proportional to the
coupling to the environment $\Gamma$, and (2)
the occupation number before the avalanche has no direct consequence on
the strength of $E_{\rm av}$. The fact that $E_{\rm av}\propto\Gamma$
indicates that the avalanche arises from the formation of off-edge states
established during the timescale set by the dephasing time $\Gamma^{-1}$. That
is, with a long-dephasing time ($\Gamma\to 0$), the multi-phonon replica
becomes more robust with a smaller electric field. The
second observation directly points to the quantum nature of the
transition. It is highly counter-intuitive that the initial occupation,
which reflects the thermal occupation of the conduction band via
the line-broadening $\Gamma$, goes against the avalanche transition. As
will be argued shortly about the gap dependence of the
avalanche, the avalanche only requires the existence of the particle
source, but not the proximity of the particle reservoirs.
We confirmed numerically that the avalanche occurs in square and cubic
lattices.

\begin{figure}
\begin{center}
\rotatebox{0}{\resizebox{3.5in}{!}{\includegraphics{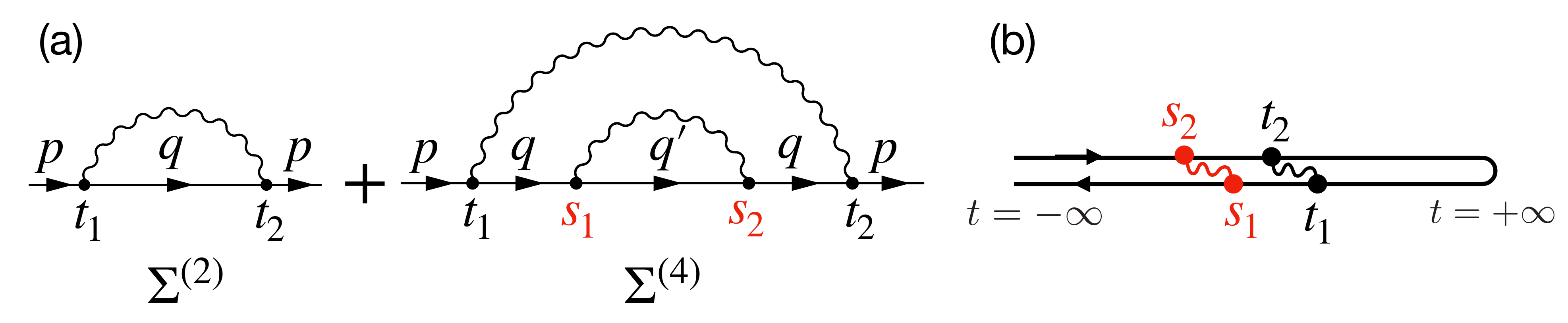}}}
\caption{(a) Lowest-order self-energy to electron by electron-phonon
coupling. Electron (solid line) emits/absorbs a phonon (wiggly line) in
the scattering. (b) The next-order self-energy showing two-phonon
process. The integral is performed over the internal (red) Keldysh times $s_1$
and $s_2$. (c) Out of the 12 possible arrangements of $(s_1,s_2)$ on the
Keldysh contour, the dominant contribution comes with $s_{1,2}<t_{1,2}$ on each
contour, as shown.
}
\label{fig2}
\end{center}
\end{figure}

Now, we identify the condition for the abrupt increase of electron
occupation in the band. The electron occupation is directly obtained
from the lesser Green's function (GF),
$
n_{\rm ex}=-i\int G^<_p(t,t)\frac{dp}{2\pi},
$
with the lesser GF defined as
$
G^<_p(t_2,t_1)=i\langle c^\dagger_p(t_1)c_p(t_2)\rangle,
$
with the Fourier transformed fermion variable $c_p=\int
e^{-ipx}\psi(x)dx$.  The enhancement of occupation results from the
lesser self-energy $\Sigma^<$, symbolically through $G^< = G^R\Sigma^<
G^A$. The details of the calculation are given in the SM.
FIG.~\ref{fig2}(a) and (b) represent the two lowest-order self-energies
due to one-phonon and two-phonon emission, respectively, and we look for
the condition that these processes lead to comparable magnitude so that
we expect an infinite summation of these 'rainbow' diagrams leads to an
occupation avalanche.

The lowest-order self-energy $\Sigma^{(2),<}_p(t_2,t_1)$,
FIG.~\ref{fig2}{a), can be written~\cite{khan_wilkins} as
\begin{equation}
\Sigma^{(2),<}_p(t_2,t_1)=ig_{\rm
ep}^2\int\frac{dq}{2\pi}D^<_0(t_2,t_1)G^<_{0q}(t_2,t_1),
\end{equation}
where $D^<_0(t_2,t_1)$ is the standard Keldysh GF for
phonon, and $G^<_{0q}(t_2,t_1)$ for non-interacting electron with
momentum $q$. The electronic lesser GF is given as
\begin{equation}
G^<_{0p}(t_2,t_1)=in_{\rm ex}(p)e^{-\Gamma|t_2-t_1|}U_p(t_2,t_1),
\end{equation}
with an unspecified initial occupation $n_{\rm ex}(p)$ and the time-evolution
factor of a free electron
\begin{eqnarray}
U_p(t_2,t_1)&=&\exp\left[-i\int_{t_1}^{t_2}\left(\frac{(p+Es)^2}{2m}+\Delta\right)ds\right]
\\
&=&
\exp\left[-i\left(
\frac{(p+ET)^2}{2m}+\Delta\right) t-\frac{iE^2t^3}{24m}
\right], \nonumber
\end{eqnarray}
with the average time $T=\frac12(t_2+t_1)$ and the relative time
$t=t_2-t_1$. We note that physical observables are
gauge-independent~\cite{aron2012prl,onoda_2006,han_prb2013} and
the mechanical momentum $\bar{p}=p+ET$ appears in the manner as above. Assuming that
the occupation of the conduction band is only at $\bar{p}\approx 0$
up to the onset of the avalanche and that the bath temperature $T$ is much smaller
than the phonon energy $\omega_0$, we obtain
\begin{equation}
\Sigma^{(2),<}_p(t_2,t_1)\approx
\frac{in_{\rm ex}g_{\rm
ep}^2}{2\omega_0}e^{-\Gamma|t|-i[(\Delta-\omega_0)t+E^2t^3/24m]}.
\end{equation}
Note that the bandedge $\Delta$ is shifted down by $\omega_0$ due to the
phonon-emission.

The next-order self-energy, while we only look at the nested diagram,
can be quite formidable due to the 12 different arrangements
of the internal Keldysh times $s_1$ and $s_2$, see FIG.~\ref{fig2}(b).
However, using the fact that $|G^<| \ll |G^>|$ in the dilute limit, the only
dominant time-ordering is as shown in (b), $-\infty<s_{1,2}<t_{1,2}$ in
the backward/forward Keldysh time-contour, respectively. As detailed in
SM, $\Sigma^{(4),<}_p(t)$ is approximated as
\begin{eqnarray}
\Sigma^{(4),<}_p(t)
& \approx 
& -\frac{n_{\rm ex}mg_{\rm ep}^4}{(2\omega_0)^2}e^{-i[(\Delta-2\omega_0)t+E^2t^3/24m]}
\nonumber \\
& & \quad\times
\int\frac{dq}{2\pi}\int
ds\frac{e^{i(q^2/2m+\omega_0)s}}{qE(t+s)+2im\Gamma}.
\end{eqnarray}
In the above approximation, it is crucial that the dephasing rate $\Gamma$
and the resulting electric field $E$ are much smaller than other energy
scales such as the phonon frequency $\omega_0$ and the kinetic energy.
We define the figure of merit value $\lambda$ for the enhancement of
multi-phonon effect as
$\Sigma^{(4),<}_p(0)/\Sigma^{(2),<}_p(0)$, and
\begin{equation}
\lambda\approx \frac{img_{\rm ep}^2}{2\omega_0}\int\frac{dq}{2\pi}\int
ds\frac{e^{i(q^2/2m+\omega_0)s}}{qEs+2im\Gamma}.
\label{lambda}
\end{equation}
This integral is readily expressed in terms of the modified Bessel
function $K_0(x)$, and we arrive at the avalanche condition $\lambda=1$
as
\begin{equation}
1=\frac{mg_{\rm ep}^2}{\omega_0
E}K_0\left(\frac{2\Gamma\sqrt{2m\omega_0}}{E}\right),
\label{criterion}
\end{equation}
which is one of our main results.

We turn our attention to the integral, Eq.~(\ref{lambda}), which is
quite revealing. In the $E=0$ limit, the $s$-integral gives
$\delta(q^2/2m+\omega_0)=0$ due to the absence of target electronic
states below the band edge, as depicted in FIG.~\ref{fig1}(a). Therefore, in
the equilibrium limit, there is no enhancement for an avalanche. At a
finite $E$, the pole at $qs=-2im\Gamma/E$ results in a dominant
contribution while smearing the energy conservation due to the
time-dependent Hamiltonian, and leads to a strong enhancement for an
avalanche. Further discussions on this issue are given in SM.

Using the asymptotic relation of $K_0(x)$ for large $x$, we can express
the solution in the small avalanche field $E_{\rm av}$ limit as
\begin{equation}
E_{\rm av}\approx
\frac{2\Gamma\sqrt{2m\omega_0}}{\ln\left(\sqrt{\frac{\pi}{32}}\beta
\right)},
\end{equation}
with the avalanche parameter defined as
\begin{equation}
\beta = \frac{g_{\rm
ep}^2}{\Gamma}\left(\frac{2m}{\omega_0^3}\right)^{1/2}.
\label{beta}
\end{equation}

\begin{figure}
\begin{center}
\rotatebox{0}{\resizebox{3.5in}{!}{\includegraphics{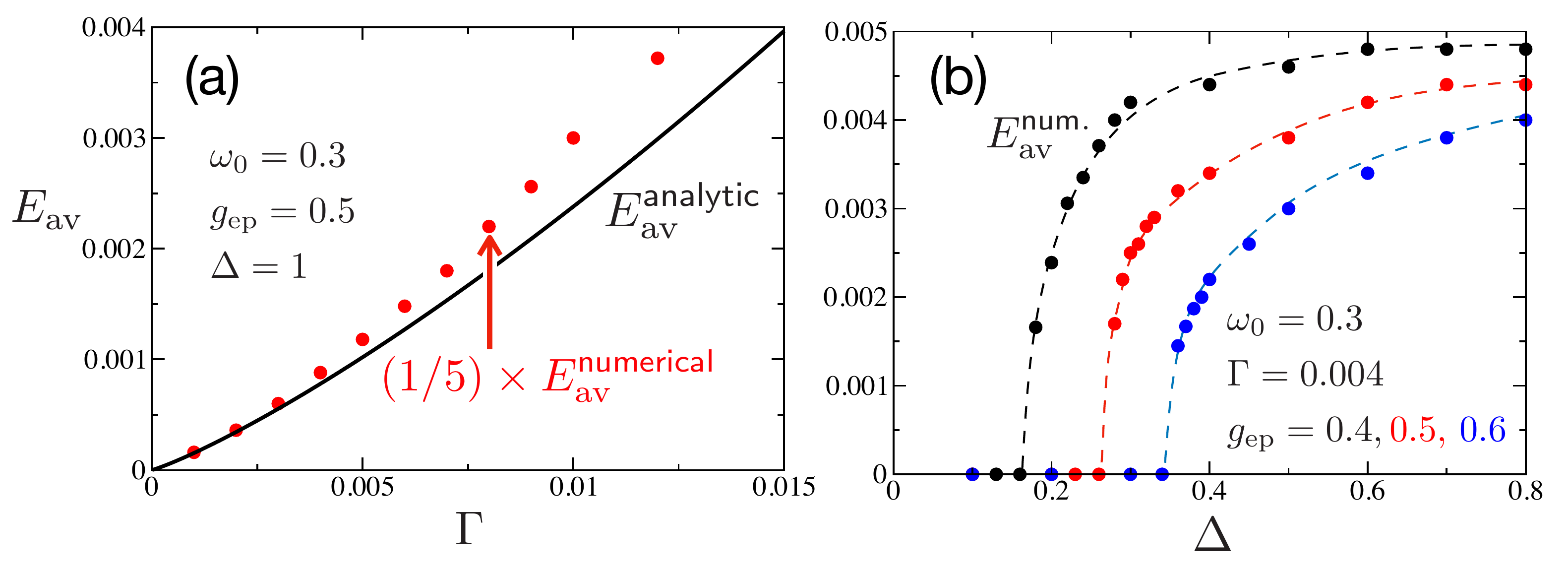}}}
\caption{
(a) Avalanche field $E_{\rm av}$ versus dephasing rate $\Gamma$,
numerical (data points) and analytic (solid line) results. $E_{\rm av}\to
0$ as $\Gamma\to 0$ almost linearly. (b) Saturation of $E_{\rm av}$ with
large $\Delta$. The saturation shows that the avalanche is due to the
fundamental instability of conduction band itself, not due to the
proximity to the particle reservoir. The initial absence of $E_{\rm av}$ at small $\Delta$ is due
to shift of gap by the electron-phonon coupling. The dashed lined are guide to the eye.
}
\label{fig3}
\end{center}
\end{figure}

In the following, we will compare the avalanche condition (\ref{criterion}) against
numerical results. Two of the most fundamental aspects of the quantum
avalanche are displayed in FIG.~\ref{fig3}. In (a), almost linear
dependence between $E_{\rm av}$ and $\Gamma$ illustrates te fact that the $\Gamma=0$
limit is fundamentally singular in the nonequilibrium
limit~\cite{han_avalanche}, and that the
dephasing is crucial in understanding nonequilibrium steady-state. 

The saturation of $E_{\rm av}$ for large $\Delta$ in FIG.~\ref{fig3}(b)
is quite surprising. The weak dependence of $E_{\rm av}$ on large $\Delta$
suggests that the avalanche is not a function of the proximity of the
band to the particle source, but rather a fundamental instability of the
conduction band itself once the particle source is accessible. This observation is consistent with
FIG.~\ref{fig1}(c), in which the avalanche occurred earlier with lower
initial occupations in the band. In the
analytic argument, it is a quite natural conclusion since
Eq.~(\ref{criterion}) is a result of integration between fermion GFs
where the energy difference enters and the gap $\Delta$ dependence drops
out. The following discussions resulting from Eq.~(\ref{criterion})
correspond to the large $\Delta$ limit. We
caution here that $E_{\rm av}$ saturates only with the model parameter
$\Delta$ and, in physical systems, the gap dependence could come back
indirectly since the bandgap is roughly proportional to the phonon energy~\cite{cardona}.

\begin{figure}
\begin{center}
\rotatebox{0}{\resizebox{3.5in}{!}{\includegraphics{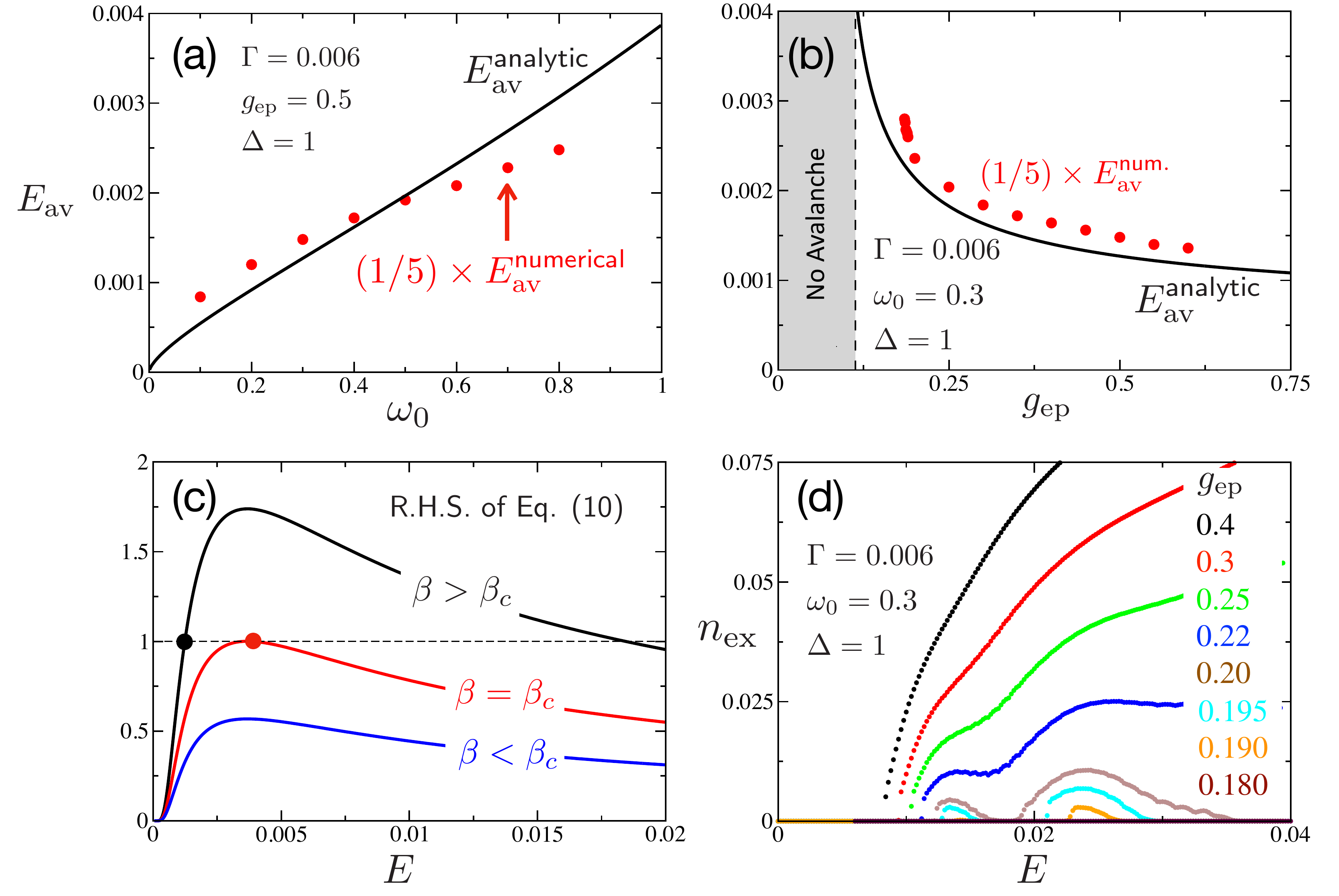}}}
\caption{
(a) Avalanche field $E_{\rm av}$ versus phonon frequency $\omega_0$.
The electron replica generation 
is inversely proportional to $\omega_0$, thus needing a larger $E$-field to
generate high phonon energy.
(b) $E_{\rm av}$ versus electron-phonon coupling
$g_{\rm ep}$. Not only the inverse dependence on $g_{\rm ep}$ but also
the sharp increase at a threshold $g_{\rm ep}$ is well agreed between the
numerical and analytic results. (c) Graphs for
criterion Eq.~(\ref{criterion}), as $g_{\rm ep}$ is varied. Solutions
(dots) cease to exist for $\beta < \beta_c$. (d) Numerical
results of $n_{\rm ex}$ vs $E$-field as $g_{\rm ep}$ is varied. As
$g_{\rm ep}$ is approached the threshold value, the avalanche becomes
reentrant with the size of domes eventually diminishing to zero at the
threshold $g_{\rm ep}$.
}
\label{fig4}
\end{center}
\end{figure}

The dependence on the phonon parameters $\omega_0$ and $g_{\rm ep}$ are
shown in FIG.~\ref{fig4}(a) and (b). The monotonic dependence is
expected since a larger phonon energy requires a stronger $E$-field. It is
still remarkable that the curvature change agrees between the analytic
and numerical results. The coupling constant dependence in (b) is
impressive. The inverse dependence is expected since a weak coupling
would need a higher field to generate an avalanche. What is interesting
is that the sharp increase occurs at a finite value of $g_{\rm ep}$.

The threshold behavior of $g_{\rm ep}$ can be understood analytically.
As shown in FIG.~\ref{fig4}(c), the R.H.S. of Eq.~(\ref{criterion}) is
a non-monotonic function. As $g_{\rm ep}$ decreases the maximum of the
R.H.S. hits 1, after which there does not exist an avalanche solution
anymore. By parametrizing $x=2\Gamma\sqrt{2m\omega_0}/E$, this
condition amounts to $[xK_0(x)]'_{x=x_0}=0$ at $x_0=0.595047$.
Eq.~(\ref{criterion}) can be rewritten as $xK_0(x)=4/\beta$ with
Eq.~(\ref{beta}). Therefore, the critical condition for the existence
of an avalanche becomes
\begin{equation}
\left.\frac{g_{\rm
ep}^2}{\Gamma}\left(\frac{2m}{\omega_0^3}\right)^{1/2}\right|_c
=\frac{4}{x_0K_0(x_0)}=8.574=\beta_c,
\label{phase}
\end{equation}
demonstrating the competition between the dephasing and the avalanching
mechanism.


Numerical solutions predict peculiar reentrant behaviors near the
threshold. As shown in FIG.~\ref{fig4}(d), $E_{\rm av}$ values increase
as $g_{\rm ep}$ is reduced. The $n_{\rm ex}$ curves after the avalanche
do not simply collapse to zero as $g_{\rm ep}$ is reduced, but they
develop reentrant avalanche domes before they reach $\beta=\beta_c$. We
speculate that after an avalanche the electrons get excited strongly and
the additional dephasing due to the charge-fluctuation mitigates the
avalanche leading to the dome behavior. We emphasize that this most
elementary nonequilibrium interacting model presents us with very rich
physics. The inclusion of the phonon self-energy did not change the
reentrant behavior.

\begin{figure}
\begin{center}
\rotatebox{0}{\resizebox{3.5in}{!}{\includegraphics{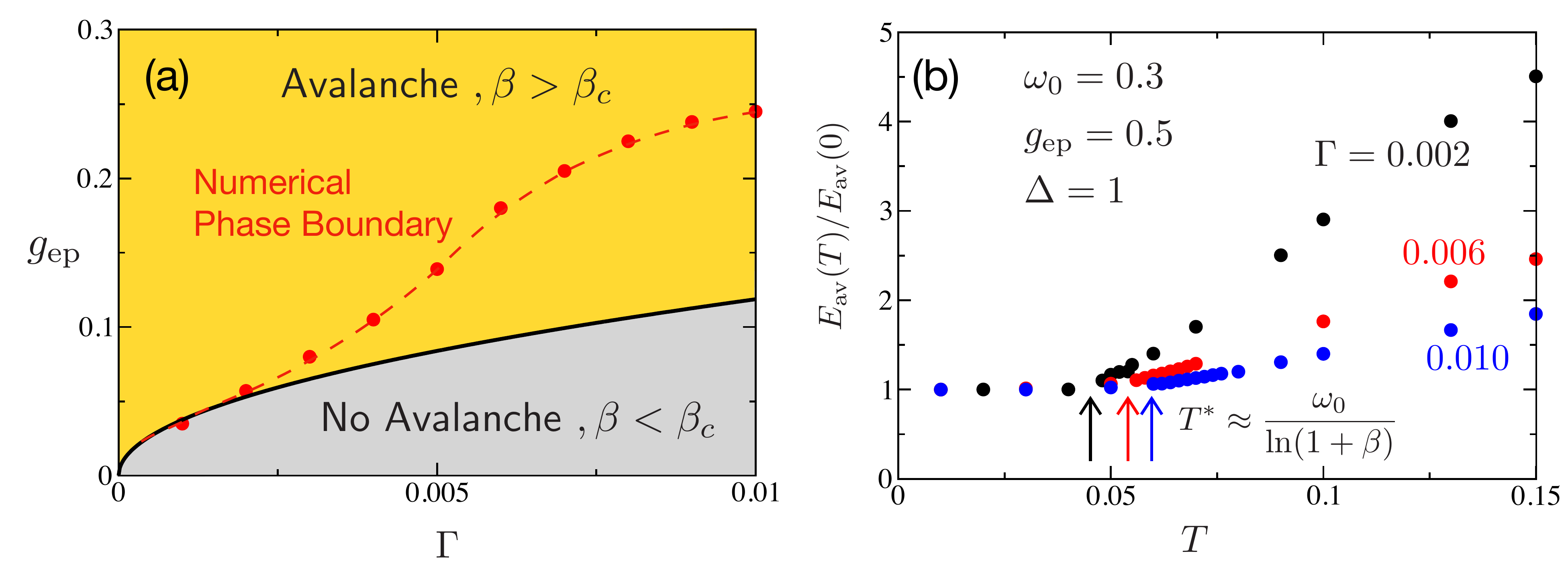}}}
\caption{
(a) $\Gamma$-$g_{\rm ep}$ phase diagram for avalanche. The competition
between the electron-phonon coupling and the dephasing results in the
phase boundary line $\beta=\beta_c$ ($g_{\rm ep}\propto\sqrt{\Gamma}$). $\beta > \beta_c$ supports
avalanche phase. $\Delta=1$ and $\omega_0=0.3$. (b) Finite temperature behavior of $E_{\rm av}$. The
increasing $E_{\rm av}(T)$ behavior is a direct evidence of a non-thermal
mechanism. The analytic prediction of the temperature $T^*$ at the onset
(arrows) agrees well with the numerical $E_{\rm av}(T)$.
}
\label{fig5}
\end{center}
\end{figure}

The competition between the avalanching mechanism and
the dephasing is best summarized in the $g_{\rm ep}$-$\Gamma$ phase
diagram for the existence of an avalanche [see FIG.~\ref{fig5}(a)]. 
According to Eq.~(\ref{phase}),
the line $\beta=\beta_c$ (or $g_{\rm ep}\propto\sqrt{\Gamma}$) divides
the phase with avalanche ($\beta>\beta_c$) and that without avalanche
($\beta<\beta_c$).

Finally, we discuss the temperature dependence of the avalanche in
FIG.~\ref{fig5}(b). In the resistive switching literature, the
conventional understanding is that the switching field of the
insulator-to-metal transition decreases with increasing temperature
since the order in solids softens with temperature~\cite{maki1986}. The quantum
avalanche~\cite{han_avalanche} mechanism has predicted the opposite
trend to this thermal scenario.  This is due to the thermal dephasing
counteracting the avalanching mechanism. Therefore, understanding the
numerical temperature dependence of $E_{\rm av}$ is crucial. As shown in
(b), the temperature $T^*$ at which the $E_{\rm av}$ changes
significantly is at $T^*\approx 0.17\omega_0$. At this temperature, the
Bose-Einstein function $n_b(\omega_0)=(e^{\omega_0/T^*}-1)^{-1}$ is less
than 0.005, which suggests that the mechanism may not be
straightforward.

The numerical temperature dependence is resolved by introducing the additional
dephasing due to the electron-phonon coupling. The dephasing is
evaluated from the retarded self-energy $\Sigma^R(\omega)$ at the
bandedge $\omega=\Delta$. Since the $\Sigma^>$ contribution dominates
$\Sigma^R$, we have in the small $E$-field limit
\begin{equation}
-{\rm Im}\Sigma^R(\Delta)\approx \frac{mg_{\rm
ep}^2}{2\omega_0\sqrt{2m\omega_0}}n_b(\omega_0)
=\frac14\Gamma\beta n_b(\omega_0).
\end{equation}
By replacing $\Gamma$ in Eq.~(\ref{criterion}) by the effective
dephasing $\Gamma+|{\rm Im}\Sigma^R(\Delta)|$, we define the activation
temperature $T^*$ at $|{\rm Im}\Sigma^R(\Delta)|\approx \frac14\Gamma$
and obtain
\begin{equation}
T^* =\frac{\omega_0}{\ln(1+\beta)}.
\end{equation}
The position for $T^*$, marked by arrows in FIG.~\ref{fig5}(b), agrees
well with the numerical results.

In conclusion, we have proposed the avalanche instability of a
conduction band under an electric field, as a general mechanism for
a nonequilibrium quantum criticality. Analytical theory, with a
comprehensive agreement with numerical confirmation, demonstrates the
quantum origin of the avalanche and presents a step toward
understanding the quantum nature of the nonequilibrium phase transition.
Further studies are necessary to test the ubiquity of the mechanism under
various dephasing mechanisms.


\begin{thebibliography}{31}%
\makeatletter
\providecommand \@ifxundefined [1]{%
 \@ifx{#1\undefined}
}%
\providecommand \@ifnum [1]{%
 \ifnum #1\expandafter \@firstoftwo
 \else \expandafter \@secondoftwo
 \fi
}%
\providecommand \@ifx [1]{%
 \ifx #1\expandafter \@firstoftwo
 \else \expandafter \@secondoftwo
 \fi
}%
\providecommand \natexlab [1]{#1}%
\providecommand \enquote  [1]{``#1''}%
\providecommand \bibnamefont  [1]{#1}%
\providecommand \bibfnamefont [1]{#1}%
\providecommand \citenamefont [1]{#1}%
\providecommand \href@noop [0]{\@secondoftwo}%
\providecommand \href [0]{\begingroup \@sanitize@url \@href}%
\providecommand \@href[1]{\@@startlink{#1}\@@href}%
\providecommand \@@href[1]{\endgroup#1\@@endlink}%
\providecommand \@sanitize@url [0]{\catcode `\\12\catcode `\$12\catcode
  `\&12\catcode `\#12\catcode `\^12\catcode `\_12\catcode `\%12\relax}%
\providecommand \@@startlink[1]{}%
\providecommand \@@endlink[0]{}%
\providecommand \url  [0]{\begingroup\@sanitize@url \@url }%
\providecommand \@url [1]{\endgroup\@href {#1}{\urlprefix }}%
\providecommand \urlprefix  [0]{URL }%
\providecommand \Eprint [0]{\href }%
\providecommand \doibase [0]{http://dx.doi.org/}%
\providecommand \selectlanguage [0]{\@gobble}%
\providecommand \bibinfo  [0]{\@secondoftwo}%
\providecommand \bibfield  [0]{\@secondoftwo}%
\providecommand \translation [1]{[#1]}%
\providecommand \BibitemOpen [0]{}%
\providecommand \bibitemStop [0]{}%
\providecommand \bibitemNoStop [0]{.\EOS\space}%
\providecommand \EOS [0]{\spacefactor3000\relax}%
\providecommand \BibitemShut  [1]{\csname bibitem#1\endcsname}%
\let\auto@bib@innerbib\@empty
\bibitem [{\citenamefont {Bardeen}(1990)}]{bardeenPT}%
  \BibitemOpen
  \bibfield  {author} {\bibinfo {author} {\bibfnamefont {J.}~\bibnamefont
  {Bardeen}},\ }\href@noop {} {\bibfield  {journal} {\bibinfo  {journal}
  {Physics Today}\ ,\ \bibinfo {pages} {25}} (\bibinfo {year}
  {1990})}\BibitemShut {NoStop}%
\bibitem [{\citenamefont {Bardeen}(1989)}]{bardeen1989}%
  \BibitemOpen
  \bibfield  {author} {\bibinfo {author} {\bibfnamefont {J.}~\bibnamefont
  {Bardeen}},\ }\href {\doibase 10.1103/PhysRevB.39.3528} {\bibfield  {journal}
  {\bibinfo  {journal} {Phys. Rev. B}\ }\textbf {\bibinfo {volume} {39}},\
  \bibinfo {pages} {3528} (\bibinfo {year} {1989})},\ \bibinfo {note}
  {publisher: American Physical Society}\BibitemShut {NoStop}%
\bibitem [{\citenamefont {Ridley}(1963)}]{Ridley}%
  \BibitemOpen
  \bibfield  {author} {\bibinfo {author} {\bibfnamefont {B.~K.}\ \bibnamefont
  {Ridley}},\ }\href@noop {} {\bibfield  {journal} {\bibinfo  {journal} {Proc.
  Phys. Soc.}\ }\textbf {\bibinfo {volume} {82}},\ \bibinfo {pages} {954}
  (\bibinfo {year} {1963})}\BibitemShut {NoStop}%
\bibitem [{\citenamefont {Thorne}(2005)}]{thornehistory}%
  \BibitemOpen
  \bibfield  {author} {\bibinfo {author} {\bibfnamefont {R.~E.}\ \bibnamefont
  {Thorne}},\ }\href {\doibase 10.1051/jp4:2005131020} {\bibfield  {journal}
  {\bibinfo  {journal} {Journal de Physique IV (Proceedings)}\ }\textbf
  {\bibinfo {volume} {131}},\ \bibinfo {pages} {89} (\bibinfo {year}
  {2005})}\BibitemShut {NoStop}%
\bibitem [{\citenamefont {Janod}\ \emph {et~al.}(2015)\citenamefont {Janod},
  \citenamefont {Tranchant}, \citenamefont {Corraze}, \citenamefont
  {Querr{\'e}}, \citenamefont {Stoliar}, \citenamefont {Rozenberg},
  \citenamefont {Cren}, \citenamefont {Roditchev}, \citenamefont {Phuoc},
  \citenamefont {Besland},\ and\ \citenamefont {Cario}}]{janod}%
  \BibitemOpen
  \bibfield  {author} {\bibinfo {author} {\bibfnamefont {E.}~\bibnamefont
  {Janod}}, \bibinfo {author} {\bibfnamefont {J.}~\bibnamefont {Tranchant}},
  \bibinfo {author} {\bibfnamefont {B.}~\bibnamefont {Corraze}}, \bibinfo
  {author} {\bibfnamefont {M.}~\bibnamefont {Querr{\'e}}}, \bibinfo {author}
  {\bibfnamefont {P.}~\bibnamefont {Stoliar}}, \bibinfo {author} {\bibfnamefont
  {M.}~\bibnamefont {Rozenberg}}, \bibinfo {author} {\bibfnamefont
  {T.}~\bibnamefont {Cren}}, \bibinfo {author} {\bibfnamefont {D.}~\bibnamefont
  {Roditchev}}, \bibinfo {author} {\bibfnamefont {V.~T.}\ \bibnamefont
  {Phuoc}}, \bibinfo {author} {\bibfnamefont {M.-P.}\ \bibnamefont {Besland}},
  \ and\ \bibinfo {author} {\bibfnamefont {L.}~\bibnamefont {Cario}},\
  }\href@noop {} {\bibfield  {journal} {\bibinfo  {journal} {Adv. Func.
  Mater.}\ }\textbf {\bibinfo {volume} {25}},\ \bibinfo {pages} {6287}
  (\bibinfo {year} {2015})}\BibitemShut {NoStop}%
\bibitem [{\citenamefont {Aoki}\ \emph {et~al.}(2014)\citenamefont {Aoki},
  \citenamefont {Tsuji}, \citenamefont {Eckstein}, \citenamefont {Kollar},
  \citenamefont {Oka},\ and\ \citenamefont {Werner}}]{aoki}%
  \BibitemOpen
  \bibfield  {author} {\bibinfo {author} {\bibfnamefont {H.}~\bibnamefont
  {Aoki}}, \bibinfo {author} {\bibfnamefont {N.}~\bibnamefont {Tsuji}},
  \bibinfo {author} {\bibfnamefont {M.}~\bibnamefont {Eckstein}}, \bibinfo
  {author} {\bibfnamefont {M.}~\bibnamefont {Kollar}}, \bibinfo {author}
  {\bibfnamefont {T.}~\bibnamefont {Oka}}, \ and\ \bibinfo {author}
  {\bibfnamefont {P.}~\bibnamefont {Werner}},\ }\href@noop {} {\bibfield
  {journal} {\bibinfo  {journal} {Rev. Mod. Phys.}\ }\textbf {\bibinfo {volume}
  {86}},\ \bibinfo {pages} {779} (\bibinfo {year} {2014})}\BibitemShut
  {NoStop}%
\bibitem [{\citenamefont {Ong}\ \emph {et~al.}(1979)\citenamefont {Ong},
  \citenamefont {Brill}, \citenamefont {Eckert}, \citenamefont {Savage},
  \citenamefont {Khanna},\ and\ \citenamefont {Somoano}}]{ong1979}%
  \BibitemOpen
  \bibfield  {author} {\bibinfo {author} {\bibfnamefont {N.~P.}\ \bibnamefont
  {Ong}}, \bibinfo {author} {\bibfnamefont {J.~W.}\ \bibnamefont {Brill}},
  \bibinfo {author} {\bibfnamefont {J.~C.}\ \bibnamefont {Eckert}}, \bibinfo
  {author} {\bibfnamefont {J.~W.}\ \bibnamefont {Savage}}, \bibinfo {author}
  {\bibfnamefont {S.~K.}\ \bibnamefont {Khanna}}, \ and\ \bibinfo {author}
  {\bibfnamefont {R.~B.}\ \bibnamefont {Somoano}},\ }\href {\doibase
  10.1103/PhysRevLett.42.811} {\bibfield  {journal} {\bibinfo  {journal} {Phys.
  Rev. Lett.}\ }\textbf {\bibinfo {volume} {42}},\ \bibinfo {pages} {811}
  (\bibinfo {year} {1979})}\BibitemShut {NoStop}%
\bibitem [{\citenamefont {Gr\"uner}(1988)}]{grunerRMP}%
  \BibitemOpen
  \bibfield  {author} {\bibinfo {author} {\bibfnamefont {G.}~\bibnamefont
  {Gr\"uner}},\ }\href {\doibase 10.1103/RevModPhys.60.1129} {\bibfield
  {journal} {\bibinfo  {journal} {Rev. Mod. Phys.}\ }\textbf {\bibinfo {volume}
  {60}},\ \bibinfo {pages} {1129} (\bibinfo {year} {1988})}\BibitemShut
  {NoStop}%
\bibitem [{\citenamefont {Zimmers}\ \emph {et~al.}(2013)\citenamefont
  {Zimmers}, \citenamefont {Aigouy}, \citenamefont {Mortier}, \citenamefont
  {Sharoni}, \citenamefont {Wang}, \citenamefont {West}, \citenamefont
  {Ramirez},\ and\ \citenamefont {Schuller}}]{zimmers}%
  \BibitemOpen
  \bibfield  {author} {\bibinfo {author} {\bibfnamefont {A.}~\bibnamefont
  {Zimmers}}, \bibinfo {author} {\bibfnamefont {L.}~\bibnamefont {Aigouy}},
  \bibinfo {author} {\bibfnamefont {M.}~\bibnamefont {Mortier}}, \bibinfo
  {author} {\bibfnamefont {A.}~\bibnamefont {Sharoni}}, \bibinfo {author}
  {\bibfnamefont {S.}~\bibnamefont {Wang}}, \bibinfo {author} {\bibfnamefont
  {K.~G.}\ \bibnamefont {West}}, \bibinfo {author} {\bibfnamefont {J.~G.}\
  \bibnamefont {Ramirez}}, \ and\ \bibinfo {author} {\bibfnamefont {I.~K.}\
  \bibnamefont {Schuller}},\ }\href@noop {} {\bibfield  {journal} {\bibinfo
  {journal} {Phys. Rev. Lett.}\ }\textbf {\bibinfo {volume} {110}},\ \bibinfo
  {pages} {056601} (\bibinfo {year} {2013})}\BibitemShut {NoStop}%
\bibitem [{\citenamefont {Zhang}\ \emph {et~al.}(2019)\citenamefont {Zhang},
  \citenamefont {McLeod}, \citenamefont {Han}, \citenamefont {Chen},
  \citenamefont {Bechtel}, \citenamefont {Yao}, \citenamefont {Gilbert~Corder},
  \citenamefont {Ciavatti}, \citenamefont {Tao}, \citenamefont {Aronson},
  \citenamefont {Carr}, \citenamefont {Martin}, \citenamefont {Sow},
  \citenamefont {Yonezawa}, \citenamefont {Nakamura}, \citenamefont {Terasaki},
  \citenamefont {Basov}, \citenamefont {Millis}, \citenamefont {Maeno},\ and\
  \citenamefont {Liu}}]{zhang_nano}%
  \BibitemOpen
  \bibfield  {author} {\bibinfo {author} {\bibfnamefont {J.}~\bibnamefont
  {Zhang}}, \bibinfo {author} {\bibfnamefont {A.~S.}\ \bibnamefont {McLeod}},
  \bibinfo {author} {\bibfnamefont {Q.}~\bibnamefont {Han}}, \bibinfo {author}
  {\bibfnamefont {X.}~\bibnamefont {Chen}}, \bibinfo {author} {\bibfnamefont
  {H.~A.}\ \bibnamefont {Bechtel}}, \bibinfo {author} {\bibfnamefont
  {Z.}~\bibnamefont {Yao}}, \bibinfo {author} {\bibfnamefont {S.~.~I.}\
  \bibnamefont {Gilbert~Corder}}, \bibinfo {author} {\bibfnamefont
  {T.}~\bibnamefont {Ciavatti}}, \bibinfo {author} {\bibfnamefont {T.~H.}\
  \bibnamefont {Tao}}, \bibinfo {author} {\bibfnamefont {M.}~\bibnamefont
  {Aronson}}, \bibinfo {author} {\bibfnamefont {G.~.~I.}\ \bibnamefont {Carr}},
  \bibinfo {author} {\bibfnamefont {M.~C.}\ \bibnamefont {Martin}}, \bibinfo
  {author} {\bibfnamefont {C.}~\bibnamefont {Sow}}, \bibinfo {author}
  {\bibfnamefont {S.}~\bibnamefont {Yonezawa}}, \bibinfo {author}
  {\bibfnamefont {F.}~\bibnamefont {Nakamura}}, \bibinfo {author}
  {\bibfnamefont {I.}~\bibnamefont {Terasaki}}, \bibinfo {author}
  {\bibfnamefont {D.~.~I.}\ \bibnamefont {Basov}}, \bibinfo {author}
  {\bibfnamefont {A.~J.}\ \bibnamefont {Millis}}, \bibinfo {author}
  {\bibfnamefont {Y.}~\bibnamefont {Maeno}}, \ and\ \bibinfo {author}
  {\bibfnamefont {M.}~\bibnamefont {Liu}},\ }\href {\doibase
  10.1103/PhysRevX.9.011032} {\bibfield  {journal} {\bibinfo  {journal} {Phys.
  Rev. X}\ }\textbf {\bibinfo {volume} {9}},\ \bibinfo {pages} {011032}
  (\bibinfo {year} {2019})}\BibitemShut {NoStop}%
\bibitem [{\citenamefont {Zener}(1932)}]{zener}%
  \BibitemOpen
  \bibfield  {author} {\bibinfo {author} {\bibfnamefont {C.}~\bibnamefont
  {Zener}},\ }\href@noop {} {\bibfield  {journal} {\bibinfo  {journal} {Proc.
  R. Soc. A}\ }\textbf {\bibinfo {volume} {137}},\ \bibinfo {pages} {696}
  (\bibinfo {year} {1932})}\BibitemShut {NoStop}%
\bibitem [{\citenamefont {Oka}\ \emph {et~al.}(2003)\citenamefont {Oka},
  \citenamefont {Arita},\ and\ \citenamefont {Aoki}}]{oka2003}%
  \BibitemOpen
  \bibfield  {author} {\bibinfo {author} {\bibfnamefont {T.}~\bibnamefont
  {Oka}}, \bibinfo {author} {\bibfnamefont {R.}~\bibnamefont {Arita}}, \ and\
  \bibinfo {author} {\bibfnamefont {H.}~\bibnamefont {Aoki}},\ }\href@noop {}
  {\bibfield  {journal} {\bibinfo  {journal} {Phys. Rev. Lett.}\ }\textbf
  {\bibinfo {volume} {91}},\ \bibinfo {pages} {66406} (\bibinfo {year}
  {2003})}\BibitemShut {NoStop}%
\bibitem [{\citenamefont {Sugimoto}\ \emph {et~al.}(2008)\citenamefont
  {Sugimoto}, \citenamefont {Onoda},\ and\ \citenamefont {Nagaosa}}]{sugimoto}%
  \BibitemOpen
  \bibfield  {author} {\bibinfo {author} {\bibfnamefont {N.}~\bibnamefont
  {Sugimoto}}, \bibinfo {author} {\bibfnamefont {S.}~\bibnamefont {Onoda}}, \
  and\ \bibinfo {author} {\bibfnamefont {N.}~\bibnamefont {Nagaosa}},\
  }\href@noop {} {\bibfield  {journal} {\bibinfo  {journal} {Phys. Rev. B}\
  }\textbf {\bibinfo {volume} {78}},\ \bibinfo {pages} {155104} (\bibinfo
  {year} {2008})}\BibitemShut {NoStop}%
\bibitem [{\citenamefont {Eckstein}\ \emph {et~al.}(2010)\citenamefont
  {Eckstein}, \citenamefont {Oka},\ and\ \citenamefont
  {Werner}}]{eckstein2010}%
  \BibitemOpen
  \bibfield  {author} {\bibinfo {author} {\bibfnamefont {M.}~\bibnamefont
  {Eckstein}}, \bibinfo {author} {\bibfnamefont {T.}~\bibnamefont {Oka}}, \
  and\ \bibinfo {author} {\bibfnamefont {P.}~\bibnamefont {Werner}},\
  }\href@noop {} {\bibfield  {journal} {\bibinfo  {journal} {Phys. Rev. Lett.}\
  }\textbf {\bibinfo {volume} {105}},\ \bibinfo {pages} {146404} (\bibinfo
  {year} {2010})}\BibitemShut {NoStop}%
\bibitem [{\citenamefont {Han}\ \emph {et~al.}(2018)\citenamefont {Han},
  \citenamefont {Li}, \citenamefont {Aron},\ and\ \citenamefont
  {Kotliar}}]{han2018}%
  \BibitemOpen
  \bibfield  {author} {\bibinfo {author} {\bibfnamefont {J.~E.}\ \bibnamefont
  {Han}}, \bibinfo {author} {\bibfnamefont {J.}~\bibnamefont {Li}}, \bibinfo
  {author} {\bibfnamefont {C.}~\bibnamefont {Aron}}, \ and\ \bibinfo {author}
  {\bibfnamefont {G.}~\bibnamefont {Kotliar}},\ }\href {\doibase
  10.1103/PhysRevB.98.035145} {\bibfield  {journal} {\bibinfo  {journal} {Phys.
  Rev. B}\ }\textbf {\bibinfo {volume} {98}},\ \bibinfo {pages} {035145}
  (\bibinfo {year} {2018})}\BibitemShut {NoStop}%
\bibitem [{\citenamefont {Han}\ \emph {et~al.}(2023)\citenamefont {Han},
  \citenamefont {Aron}, \citenamefont {Chen}, \citenamefont {Mansaray},
  \citenamefont {Han}, \citenamefont {Kim}, \citenamefont {Randle},\ and\
  \citenamefont {Bird}}]{han_avalanche}%
  \BibitemOpen
  \bibfield  {author} {\bibinfo {author} {\bibfnamefont {J.~E.}\ \bibnamefont
  {Han}}, \bibinfo {author} {\bibfnamefont {C.}~\bibnamefont {Aron}}, \bibinfo
  {author} {\bibfnamefont {X.}~\bibnamefont {Chen}}, \bibinfo {author}
  {\bibfnamefont {I.}~\bibnamefont {Mansaray}}, \bibinfo {author}
  {\bibfnamefont {J.-H.}\ \bibnamefont {Han}}, \bibinfo {author} {\bibfnamefont
  {K.-S.}\ \bibnamefont {Kim}}, \bibinfo {author} {\bibfnamefont
  {M.}~\bibnamefont {Randle}}, \ and\ \bibinfo {author} {\bibfnamefont {J.~P.}\
  \bibnamefont {Bird}},\ }\href {\doibase 10.1038/s41467-023-38557-8}
  {\bibfield  {journal} {\bibinfo  {journal} {Nat. Comm.}\ }\textbf {\bibinfo
  {volume} {14}},\ \bibinfo {pages} {2936} (\bibinfo {year}
  {2023})}\BibitemShut {NoStop}%
\bibitem [{\citenamefont {Khan}\ \emph {et~al.}(1987)\citenamefont {Khan},
  \citenamefont {Davies},\ and\ \citenamefont {Wilkins}}]{khan_wilkins}%
  \BibitemOpen
  \bibfield  {author} {\bibinfo {author} {\bibfnamefont {F.~S.}\ \bibnamefont
  {Khan}}, \bibinfo {author} {\bibfnamefont {J.~H.}\ \bibnamefont {Davies}}, \
  and\ \bibinfo {author} {\bibfnamefont {J.~W.}\ \bibnamefont {Wilkins}},\
  }\href@noop {} {\bibfield  {journal} {\bibinfo  {journal} {Phys. Rev. B}\
  }\textbf {\bibinfo {volume} {36}},\ \bibinfo {pages} {2578} (\bibinfo {year}
  {1987})}\BibitemShut {NoStop}%
\bibitem [{\citenamefont {Kemper}\ \emph {et~al.}(2017)\citenamefont {Kemper},
  \citenamefont {Sentef}, \citenamefont {Moritz}, \citenamefont {Devereaux},\
  and\ \citenamefont {Freericks}}]{kemper}%
  \BibitemOpen
  \bibfield  {author} {\bibinfo {author} {\bibfnamefont {A.~F.}\ \bibnamefont
  {Kemper}}, \bibinfo {author} {\bibfnamefont {M.~A.}\ \bibnamefont {Sentef}},
  \bibinfo {author} {\bibfnamefont {B.}~\bibnamefont {Moritz}}, \bibinfo
  {author} {\bibfnamefont {T.~P.}\ \bibnamefont {Devereaux}}, \ and\ \bibinfo
  {author} {\bibfnamefont {J.~K.}\ \bibnamefont {Freericks}},\ }\href@noop {}
  {\bibfield  {journal} {\bibinfo  {journal} {Annalen der Physik}\ }\textbf
  {\bibinfo {volume} {529}},\ \bibinfo {pages} {1600235} (\bibinfo {year}
  {2017})}\BibitemShut {NoStop}%
\bibitem [{\citenamefont {Nathawat}\ \emph {et~al.}(2023)\citenamefont
  {Nathawat}, \citenamefont {Mansaray}, \citenamefont {Sakanashi},
  \citenamefont {Wada}, \citenamefont {Randle}, \citenamefont {Yin},
  \citenamefont {He}, \citenamefont {Arabchigavkani}, \citenamefont {Dixit},
  \citenamefont {Barut}, \citenamefont {Zhao}, \citenamefont {Ramamoorthy},
  \citenamefont {Somphonsane}, \citenamefont {Kim}, \citenamefont {Watanabe},
  \citenamefont {Taniguchi}, \citenamefont {Aoki}, \citenamefont {Han},\ and\
  \citenamefont {Bird}}]{moire2023}%
  \BibitemOpen
  \bibfield  {author} {\bibinfo {author} {\bibfnamefont {J.}~\bibnamefont
  {Nathawat}}, \bibinfo {author} {\bibfnamefont {I.}~\bibnamefont {Mansaray}},
  \bibinfo {author} {\bibfnamefont {K.}~\bibnamefont {Sakanashi}}, \bibinfo
  {author} {\bibfnamefont {N.}~\bibnamefont {Wada}}, \bibinfo {author}
  {\bibfnamefont {M.~D.}\ \bibnamefont {Randle}}, \bibinfo {author}
  {\bibfnamefont {S.}~\bibnamefont {Yin}}, \bibinfo {author} {\bibfnamefont
  {K.}~\bibnamefont {He}}, \bibinfo {author} {\bibfnamefont {N.}~\bibnamefont
  {Arabchigavkani}}, \bibinfo {author} {\bibfnamefont {R.}~\bibnamefont
  {Dixit}}, \bibinfo {author} {\bibfnamefont {B.}~\bibnamefont {Barut}},
  \bibinfo {author} {\bibfnamefont {M.}~\bibnamefont {Zhao}}, \bibinfo {author}
  {\bibfnamefont {H.}~\bibnamefont {Ramamoorthy}}, \bibinfo {author}
  {\bibfnamefont {R.}~\bibnamefont {Somphonsane}}, \bibinfo {author}
  {\bibfnamefont {G.-H.}\ \bibnamefont {Kim}}, \bibinfo {author} {\bibfnamefont
  {K.}~\bibnamefont {Watanabe}}, \bibinfo {author} {\bibfnamefont
  {T.}~\bibnamefont {Taniguchi}}, \bibinfo {author} {\bibfnamefont
  {N.}~\bibnamefont {Aoki}}, \bibinfo {author} {\bibfnamefont {J.~E.}\
  \bibnamefont {Han}}, \ and\ \bibinfo {author} {\bibfnamefont {J.~P.}\
  \bibnamefont {Bird}},\ }\href {\doibase 10.1038/s41467-023-37292-4}
  {\bibfield  {journal} {\bibinfo  {journal} {Nat. Comm.}\ }\textbf {\bibinfo
  {volume} {14}},\ \bibinfo {pages} {1507} (\bibinfo {year}
  {2023})}\BibitemShut {NoStop}%
\bibitem [{\citenamefont {Mazzocchi}\ \emph {et~al.}(2022)\citenamefont
  {Mazzocchi}, \citenamefont {Gazzaneo}, \citenamefont {Lotze},\ and\
  \citenamefont {Arrigoni}}]{mazzocchi}%
  \BibitemOpen
  \bibfield  {author} {\bibinfo {author} {\bibfnamefont {T.~M.}\ \bibnamefont
  {Mazzocchi}}, \bibinfo {author} {\bibfnamefont {P.}~\bibnamefont {Gazzaneo}},
  \bibinfo {author} {\bibfnamefont {J.}~\bibnamefont {Lotze}}, \ and\ \bibinfo
  {author} {\bibfnamefont {E.}~\bibnamefont {Arrigoni}},\ }\href {\doibase
  10.1103/PhysRevB.106.125123} {\bibfield  {journal} {\bibinfo  {journal}
  {Phys. Rev. B}\ }\textbf {\bibinfo {volume} {106}},\ \bibinfo {pages}
  {125123} (\bibinfo {year} {2022})}\BibitemShut {NoStop}%
\bibitem [{\citenamefont {Zhang}\ and\ \citenamefont
  {Chern}(2022)}]{zhang_chern}%
  \BibitemOpen
  \bibfield  {author} {\bibinfo {author} {\bibfnamefont {S.}~\bibnamefont
  {Zhang}}\ and\ \bibinfo {author} {\bibfnamefont {G.-W.}\ \bibnamefont
  {Chern}},\ }\href {http://arxiv.org/abs/2201.02194} {\  (\bibinfo {year}
  {2022})},\ \bibinfo {note} {arXiv:2201.02194 [cond-mat]}\BibitemShut
  {NoStop}%
\bibitem [{\citenamefont {Weiss}(2008)}]{weiss}%
  \BibitemOpen
  \bibfield  {author} {\bibinfo {author} {\bibfnamefont {U.}~\bibnamefont
  {Weiss}},\ }\href@noop {} {\emph {\bibinfo {title} {Quantum Dissipative
  Systems}}}\ (\bibinfo  {publisher} {World Scientific},\ \bibinfo {address}
  {London},\ \bibinfo {year} {2008})\BibitemShut {NoStop}%
\bibitem [{\citenamefont {Khurgin}\ \emph {et~al.}(2007)\citenamefont
  {Khurgin}, \citenamefont {Ding},\ and\ \citenamefont {Jena}}]{khurgin2007}%
  \BibitemOpen
  \bibfield  {author} {\bibinfo {author} {\bibfnamefont {J.}~\bibnamefont
  {Khurgin}}, \bibinfo {author} {\bibfnamefont {Y.~J.}\ \bibnamefont {Ding}}, \
  and\ \bibinfo {author} {\bibfnamefont {D.}~\bibnamefont {Jena}},\ }\href
  {\doibase 10.1063/1.2824872} {\bibfield  {journal} {\bibinfo  {journal}
  {Appl. Phys. Lett.}\ }\textbf {\bibinfo {volume} {91}},\ \bibinfo {pages}
  {252104} (\bibinfo {year} {2007})}\BibitemShut {NoStop}%
\bibitem [{\citenamefont {Keldysh}(1958)}]{franzkeldysh}%
  \BibitemOpen
  \bibfield  {author} {\bibinfo {author} {\bibfnamefont {L.~V.}\ \bibnamefont
  {Keldysh}},\ }\href@noop {} {\bibfield  {journal} {\bibinfo  {journal} {Sov.
  Phys. JETP}\ }\textbf {\bibinfo {volume} {34}},\ \bibinfo {pages} {788}
  (\bibinfo {year} {1958})}\BibitemShut {NoStop}%
\bibitem [{\citenamefont {Li}\ \emph {et~al.}(2015)\citenamefont {Li},
  \citenamefont {{Camille Aron}}, \citenamefont {Kotliar},\ and\ \citenamefont
  {Han}}]{liprl2015}%
  \BibitemOpen
  \bibfield  {author} {\bibinfo {author} {\bibfnamefont {J.}~\bibnamefont
  {Li}}, \bibinfo {author} {\bibnamefont {{Camille Aron}}}, \bibinfo {author}
  {\bibfnamefont {G.}~\bibnamefont {Kotliar}}, \ and\ \bibinfo {author}
  {\bibfnamefont {J.~E.}\ \bibnamefont {Han}},\ }\href {\doibase DOI:
  10.1103/PhysRevLett.114.226403} {\bibfield  {journal} {\bibinfo  {journal}
  {Phys. Rev. Lett.}\ }\textbf {\bibinfo {volume} {114}},\ \bibinfo {pages}
  {226403} (\bibinfo {year} {2015})}\BibitemShut {NoStop}%
\bibitem [{\citenamefont {Li}\ and\ \citenamefont {Han}(2018)}]{ligraphene}%
  \BibitemOpen
  \bibfield  {author} {\bibinfo {author} {\bibfnamefont {J.}~\bibnamefont
  {Li}}\ and\ \bibinfo {author} {\bibfnamefont {J.~E.}\ \bibnamefont {Han}},\
  }\href {\doibase 10.1103/PhysRevB.97.205412} {\bibfield  {journal} {\bibinfo
  {journal} {Phys. Rev. B}\ }\textbf {\bibinfo {volume} {97}},\ \bibinfo
  {pages} {205412} (\bibinfo {year} {2018})}\BibitemShut {NoStop}%
\bibitem [{\citenamefont {Aron}\ \emph {et~al.}(2012)\citenamefont {Aron},
  \citenamefont {Kotliar},\ and\ \citenamefont {Weber}}]{aron2012prl}%
  \BibitemOpen
  \bibfield  {author} {\bibinfo {author} {\bibfnamefont {C.}~\bibnamefont
  {Aron}}, \bibinfo {author} {\bibfnamefont {G.}~\bibnamefont {Kotliar}}, \
  and\ \bibinfo {author} {\bibfnamefont {C.}~\bibnamefont {Weber}},\
  }\href@noop {} {\bibfield  {journal} {\bibinfo  {journal} {Phys. Rev. Lett.}\
  }\textbf {\bibinfo {volume} {108}},\ \bibinfo {pages} {086401} (\bibinfo
  {year} {2012})}\BibitemShut {NoStop}%
\bibitem [{\citenamefont {Onoda}\ \emph {et~al.}(2006)\citenamefont {Onoda},
  \citenamefont {Sugimoto},\ and\ \citenamefont {Nagaosa}}]{onoda_2006}%
  \BibitemOpen
  \bibfield  {author} {\bibinfo {author} {\bibfnamefont {S.}~\bibnamefont
  {Onoda}}, \bibinfo {author} {\bibfnamefont {N.}~\bibnamefont {Sugimoto}}, \
  and\ \bibinfo {author} {\bibfnamefont {N.}~\bibnamefont {Nagaosa}},\ }\href
  {\doibase 10.1143/PTP.116.61} {\bibfield  {journal} {\bibinfo  {journal}
  {Prog. Theor. Phys.}\ }\textbf {\bibinfo {volume} {116}},\ \bibinfo {pages}
  {61} (\bibinfo {year} {2006})}\BibitemShut {NoStop}%
\bibitem [{\citenamefont {Han}(2013)}]{han_prb2013}%
  \BibitemOpen
  \bibfield  {author} {\bibinfo {author} {\bibfnamefont {J.~E.}\ \bibnamefont
  {Han}},\ }\href {\doibase 10.1103/PhysRevB.87.085119} {\bibfield  {journal}
  {\bibinfo  {journal} {Phys. Rev. B}\ }\textbf {\bibinfo {volume} {87}},\
  \bibinfo {pages} {085119} (\bibinfo {year} {2013})}\BibitemShut {NoStop}%
\bibitem [{\citenamefont {Yu}\ and\ \citenamefont {Cardona}(2010)}]{cardona}%
  \BibitemOpen
  \bibfield  {author} {\bibinfo {author} {\bibfnamefont {P.~Y.}\ \bibnamefont
  {Yu}}\ and\ \bibinfo {author} {\bibfnamefont {M.}~\bibnamefont {Cardona}},\
  }\href@noop {} {\emph {\bibinfo {title} {Fundamentals of Semiconductors}}}\
  (\bibinfo  {publisher} {Springer},\ \bibinfo {address} {Berlin},\ \bibinfo
  {year} {2010})\BibitemShut {NoStop}%
\bibitem [{\citenamefont {Maki}(1986)}]{maki1986}%
  \BibitemOpen
  \bibfield  {author} {\bibinfo {author} {\bibfnamefont {K.}~\bibnamefont
  {Maki}},\ }\href
  {http://gateway.webofknowledge.com/gateway/Gateway.cgi?GWVersion=2&SrcAuth=mekentosj&SrcApp=Papers&DestLinkType=FullRecord&DestApp=WOS&KeyUT=A1986A018700099}
  {\bibfield  {journal} {\bibinfo  {journal} {Phys. Rev. B}\ }\textbf {\bibinfo
  {volume} {33}},\ \bibinfo {pages} {2852} (\bibinfo {year}
  {1986})}\BibitemShut {NoStop}%
\end{thebibliography}

\bigskip
\centerline{\large\bf Acknowledgements}

\bigskip

Authors acknowledge the computational support from the CCR at University
at Buffalo. JEH benefited greatly from discussions with Camille Aron for
his insightful comments and encouragement, and with Ki-Seok Kim who
pointed out the importance of the multi-phonon diagrams. Helpful
discussions with Enrico Arrigoni, Gabriel Kotliar, Jonathan Bird and
Peihong Zhang are appreciated. JEH is grateful for the hospitality of
the ENS-CNRS where part of the work is completed.

\end{document}


\myfonts

\centerline{\textbf{\large\sf SUPPLEMENTARY MATERIAL
}}
\bigskip

\title{Avalanche Instability as Nonequilibrium
Quantum Criticality}

\author{Xi Chen}
\affiliation{Department of Physics, State University of New York at Buffalo, Buffalo, New York 14260, USA}
\author{Jong E. Han}
\email{jonghan@buffalo.edu}
\affiliation{Department of Physics, State University of New York at Buffalo, Buffalo, New York 14260, USA}

\maketitle
\newpage
\centerline{\bf S1. Derivation of the Avalanche Criterion, Eq.~(8)}
\bigskip

With the standard Keldysh Green's function notation~\cite{khan_wilkins}, we begin with the bare electron
propagation $G_0$ for a free electron given as
\begin{eqnarray}
G^\gtrless_{0,p}(t_2,t_1) & = & 
\mp in_\gtrless(p)
\exp\left[-\Gamma|t_2-t_1|-i\int_{t_1}^{t_2}\left(
\frac{(p+Es)^2}{2m}+\Delta\right)ds\right]
\nonumber \\
& = &
\mp in_\gtrless(p)e^{-\Gamma|t_2-t_1|}U_p(t_2,t_1)
\end{eqnarray}
with
\begin{eqnarray}
U_p(t_2,t_1) & = & 
\exp\left[-i\int_{t_1}^{t_2}\left(
\frac{(p+Es)^2}{2m}+\Delta\right)ds\right]
\\
& = &
\exp\left[-i\left(
\frac{(p+ET)^2}{2m}+\Delta\right)t-\frac{iE^2t^3}{24m}\right],
\end{eqnarray}
where $\Gamma$ is the dephasing provided by the coupling to the fermion
baths and $n_\gtrless(p)$ is the hole/electron density, $n_<(p)\ll 1$ and
$n_>(p)\approx 1-n_<(p)\approx 1$. $T$ is the average time
$T=\frac12(t_2+t_1)$ and $t$ is the relative time $t=t_2-t_1$.
While the occupation number $n_<(p)$ is to
be determined self-consistently, $n_<(p)$ is only significant for
$|p|\ll 1$. Furthermore, before the onset of an
avalanche~\cite{han_avalanche} it is given by the thermal
excitation due to the broadening provided by the hybridization $\Gamma$.
As shown in FIG.1(c), however, the avalanche is not
determined by $n_<$, which is also supported by cancellation of this factor in
the discussion below.

The bare phonon Green's function $D_0$ is defined as
\begin{equation}
D^\gtrless_0(t) = -i\langle \varphi(\pm t)\varphi(0)\rangle
= -\frac{i}{2\omega_0}\left[
n_b(\omega_0)e^{\pm i\omega_0t}+(1+n_b(\omega_0))e^{\mp i\omega_0 t}
\right],
\label{d0}
\end{equation}
with the Bose-Einstein function $n_b(\omega_0)=(e^{\omega_0/T}-1)^{-1}$.

The lowest-order electron self-energy $\Sigma^<(t_1,t_2)$ [see FIG.2(a)]
is given as
\begin{equation}
\Sigma^{(2),<}_p(t_2,t_1)=ig_{\rm
ep}^2\int\frac{dq}{2\pi}D^<_0(t_2,t_1)G^<_{0q}(t_2,t_1).
\end{equation}
We calculate the lesser self-energy since it is directly related to the
electron occupation.
In the $T=0$ limit, $n_b(\omega_0)\to 0$ and only the $e^{i\omega_0
t}$-term in Eq.~(\ref{d0}) is allowed,
\begin{equation}
\Sigma^{(2),<}_p(t_2,t_1)\approx\frac{ig_{\rm ep}^2}{2\omega_0}
\int\frac{dq}{2\pi}n_<(q)e^{-\Gamma|t|-i[(\epsilon(q+ET)-\omega_0)t+E^2t^3/24m]},
\end{equation}
with $\epsilon(q)=q^2/2m+\Delta$. Since the occupation of the conduction band is very dilute
at small momentum, we write $n_<(q)\approx 2\pi n_{\rm ex}\delta(q+ET)$ such that $n_{\rm ex}=
\int n_<(q)(dq/2\pi)$. Note that the momentum $q$ is not
gauge-independent, but the mechanical momentum $q+ET$ is. Therefore an
observable is a function of
$q+ET$~\cite{onoda_2006,aron2012prl,han_prb2013}.
Then
\begin{equation}
\Sigma^{(2),<}_p(t)\approx\frac{in_{\rm ex}g_{\rm ep}^2}{2\omega_0}
e^{-\Gamma|t|-i[(\Delta-\omega_0)t+E^2t^3/24m]}.
\end{equation}
Note that phase evolution rate from the bandedge $\Delta$ is
reduced by the phonon frequency $\omega_0$, indicating a single-phonon
emission.

\begin{figure}[h]
\rotatebox{0}{\resizebox{6.0in}{!}{\includegraphics{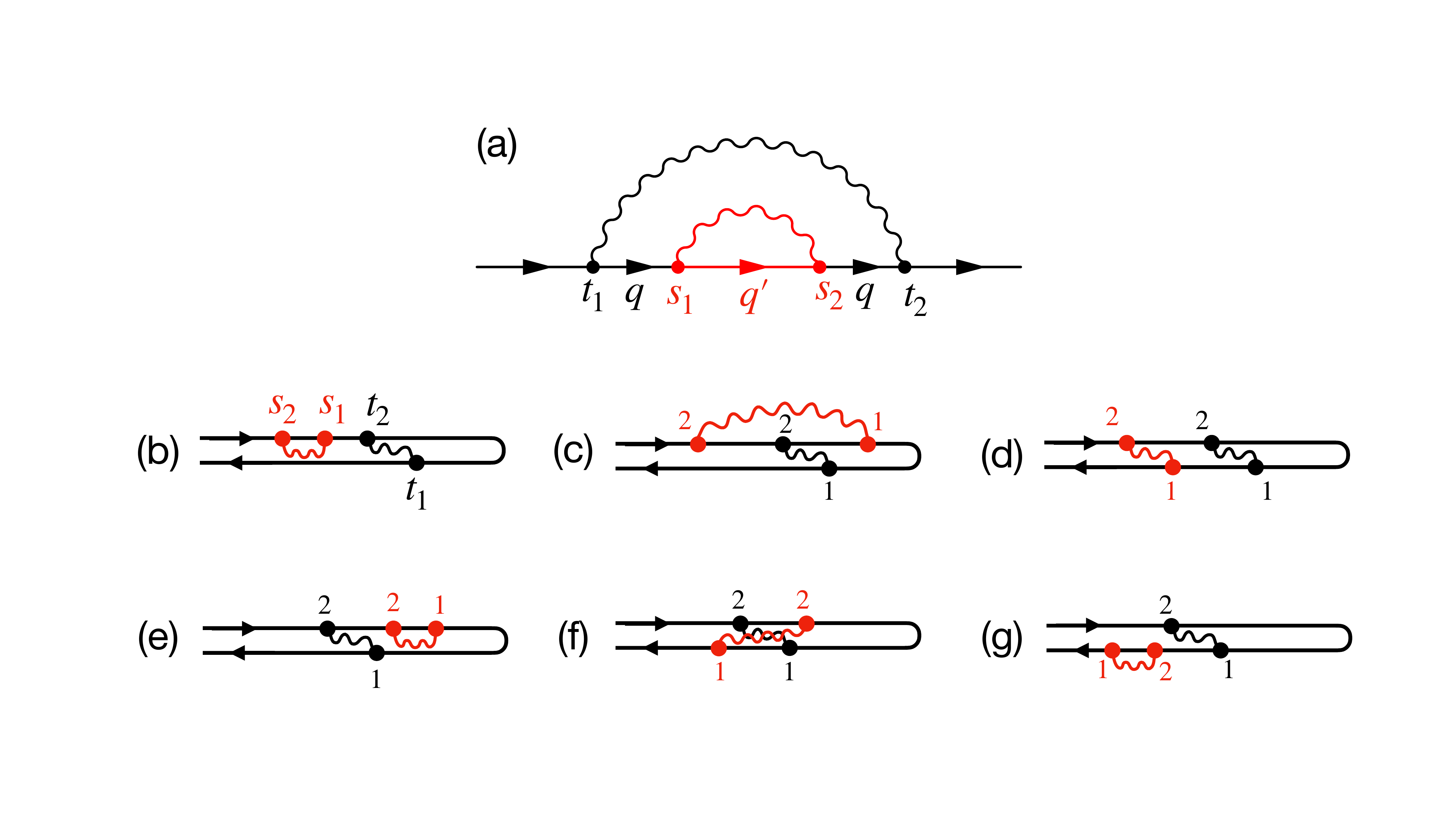}}}
\caption{(a) Fourth order electron-phonon self-energy $\Sigma^{(4)}$.
With $\Sigma^<$, the outer times $t_1$ and $t_2$ are fixed on the
Keldysh branch, and the inner times $s_1$ and $s_2$ (red) are permuted
over all possible Keldysh branch combination. (b-g) 6 different
combinations for $s_1$ and $s_2$. There are 6 more combinations with the
swapped $s_1$ and $s_2$ for each case. The order in (d) is most
dominant.
}
\label{figS1}
\end{figure}
 
We now calculate the next-order self-energy for the enhancement of the
electron-phonon coupling due to multiple-phonon process. For the diagram
considered in FIG.2(a), we express the self-energy as for time on the
Keldysh contour as
\begin{equation}
\Sigma^{(4)}(t_2,t_1)=g_{\rm ep}^4\int_K ds_1\int_K ds_2G(t_2,s_2)
D_0(t_2,t_1)G(s_2,s_1)D_0(s_2,s_1)G(s_1,t_1),
\end{equation}
where the integral is defined on the Keldysh contour. There are 12 distinct arrangements of
$(t_1,t_2)$ and $(s_1,s_2)$, given that $t_1$ is fixed on the backward Keldysh
contour and $t_2$ on the forward contour. See Fig.~\ref{figS1}. The key
to simplifying the computation is to find the leading-order in terms of $G^<$ since
it is proportional to $n_<\ll 1$. As shown in FIG.2(b-g) for different
time-ordering on the Keldysh contour, only (d) provides the
leading-order while it embodies the two-phonon emission. For instance,
with the time-ordering in (b), not only the product of Green's functions is going
to be $G^>(t_2,s_2)G^<(s_2,s_1)G^<(s_1,t_1)$ in the second-order of
$G^<$, but also it describes the process of the $(s_1,s_2)$-process that can be
absorbed into the self-energy of the one-particle GF. The same can be said
about the time-ordering of (g). With (d), the Green's function product
is $G^>(t_2,s_2)G^<(s_2,s_1)G^>(s_1,t_1)$, with $G^<$ only appearing
once. The interpretation of (d) is as follows. For the electron
occupation off the band edge, it results from sequential emissions from the past to present
(inner-loop to outer-loop) and successive lowering of electric states.

In the low-density limit, the ingap contribution to the lesser
self-energy becomes explicitly
\begin{eqnarray}
& & -i\frac{g_{\rm
ep}^4}{(2\omega_0)^2}\int\frac{dq}{2\pi}\int\frac{dq'}{2\pi}\int^{-\infty}_{t_1}ds_1
\int_{-\infty}^{t_2}ds_2 n_<(q')e^{-\Gamma(t_2-s_2+t_1-s_1+|s_2-s_1|)}
\nonumber \\
& \times &
e^{i\omega_0(s_2-s_1+t_2-t_1)}U_q(t_2,s_2)U_{q'}(s_2,s_1)U_q(s_1,t_1).
\end{eqnarray}
After substitution $s_1\to t_1+s_1$ and $s_2\to s_2+t_2$, and then by
changing the variables to the average variable $S=\frac12(s_1+s_2)$ and
the relative variable $s=s_2-s_1$, we have the integral as
\begin{eqnarray}
& & i\frac{g_{\rm
ep}^4}{(2\omega_0)^2}e^{-i[(\Delta-2\omega_0)t+E^2t^3/24m]}
\int\frac{dq}{2\pi}\int\frac{dq'}{2\pi}\int_{-\infty}^\infty ds
\int_{-\infty}^{-|s|/2}dS n_<(q')e^{\Gamma(2S-|s+t|)+i\omega_0 s}
\nonumber \\
& \times &
\exp\left[\frac{i}{2m}(q+ET)^2s-\frac{i}{2m}(q'+ET)^2(t+s)-\frac{i}{m}(q-q')E(t+s)S
\right],
\end{eqnarray}
where we used the transformation of integral
\begin{equation}
\int_{-\infty}^0 ds_1 \int_{-\infty}^0 ds_2
=\int_{-\infty}^\infty ds \int_{-\infty}^{-|s|/2}dS.
\end{equation}
As discussed above, we replace $q+ET$ and $q'+ET$ by the
gauge-independent $q$ and $q'$, respectively, and
set $q'\approx 0$ due to the factor $n_<(q')$. 
Then after performing the integral over $S$, we have
\begin{equation}
-\frac{mn_{\rm ex}g_{\rm ep}^4}{(2\omega_0)^2}e^{-i[(\Delta-2\omega_0)t+E^2t^3/24m]}
\int\frac{dq}{2\pi}\int ds
\frac{e^{-\Gamma(|s|+|s+t|)+i\omega_0 s}}{qE(t+s)+2im\Gamma}
e^{i\frac{q^2s}{2m}+i\frac{q}{2m}E(t+s)|s|}.
\end{equation}

To further simplify the expression, we make a crucial approximation
based on the parameter regime that $\frac{q^2}{2m}+\omega_0 \gg 
\Gamma\sim E$. Due to this condition, the integral's
convergence is controlled by the oscillation by
$\frac{q^2}{2m}+\omega_0$ and the exponential factors
$e^{-\Gamma(|s|+|s+t|)}$ and $e^{i\frac{q}{2m}E(t+s)|s|}$ are slowly varying
and become irrelevant. The integral is then approximated as
\begin{equation}
\Sigma^{(4),<}_p(t)\approx
-\frac{mn_{\rm ex}g_{\rm ep}^4}{(2\omega_0)^2}e^{-i[(\Delta-2\omega_0)t+E^2t^3/24m]}
\int\frac{dq}{2\pi}\int ds
\frac{e^{i(q^2/2m+\omega_0)s}}{qE(t+s)+2im\Gamma}.
\label{approximate}
\end{equation}
See the discussion in the following paragraph for further justification.
The enhancement factor $\lambda$ evaluated at $t=0$ is then
\begin{equation}
\lambda=\frac{\Sigma^{(4),<}_p(0)}{\Sigma^{(2),<}_p(0)}
\approx\frac{img_{\rm ep}^2}{2\omega_0}
\int\frac{dq}{2\pi}\int ds \frac{e^{i(q^2/2m+\omega_0)s}}{qEs+2im\Gamma}
=\frac{mg_{\rm ep}^2}{\omega_o
E}K_0\left(\frac{2\Gamma\sqrt{2m\omega_0}}{E}\right),
\label{analyticlambda}
\end{equation}
with the modified Bessel function $K_0(x)$. Note that the unspecified
parameter $n_{\rm ex}$ cancels out. We identify the avalanche at
$\lambda=1$, thus Eq.~(8) in the main text. It is noted that, since the
occupation spectra $G^<(\omega)$ is localized near
$\omega=\Delta$~\cite{han_avalanche}, its integral over $\omega$
[\textit{i.e.,} $\Sigma^<(t=0)$] is a good indicator for the avalanche.

\begin{figure}[h]
\rotatebox{0}{\resizebox{5.0in}{!}{\includegraphics{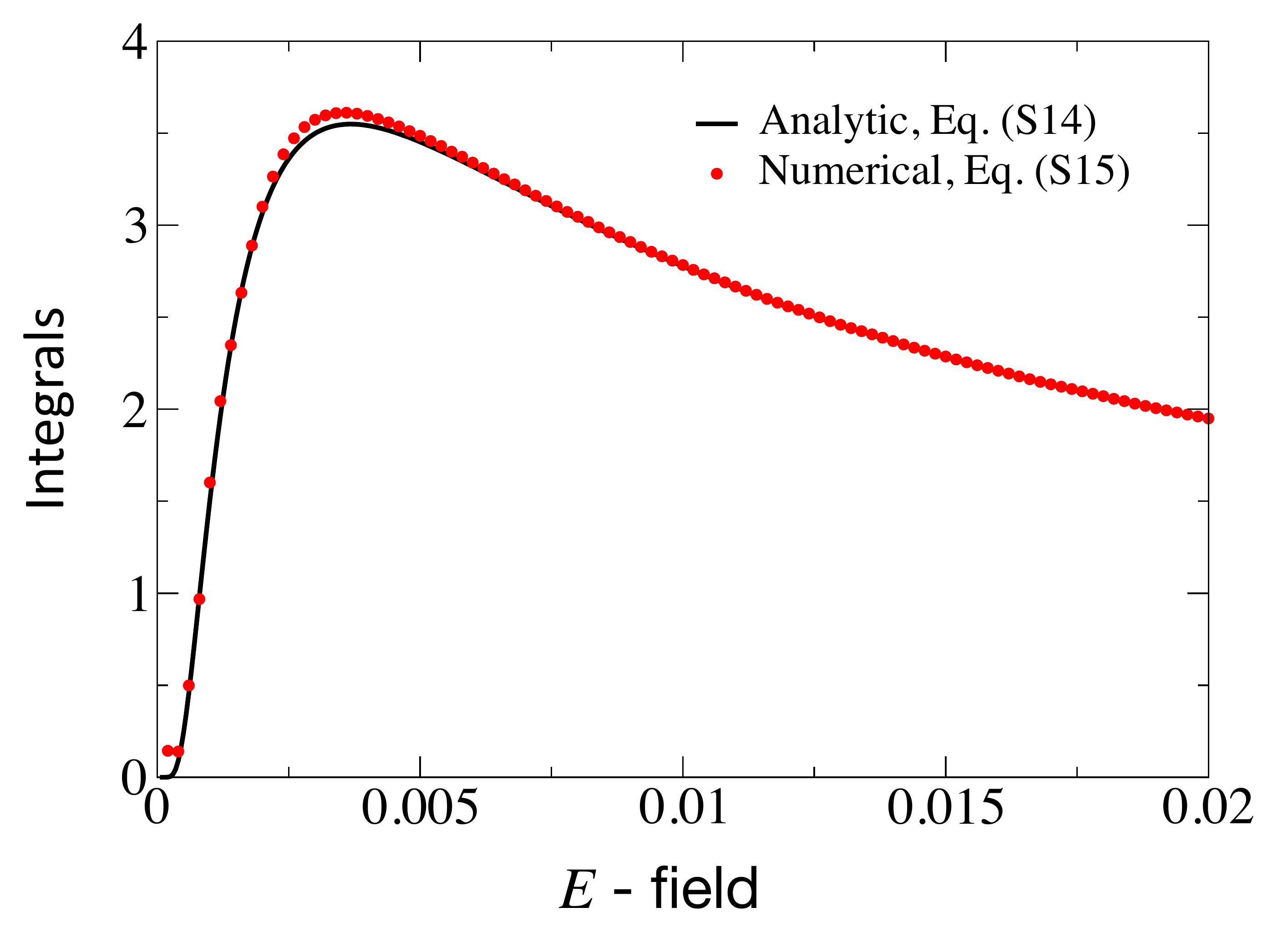}}}
\caption{Comparison of the numerical integral of Eq.~(\ref{fullint})
(circles) as a function of $E$-field 
against the analytic approximation Eq.~(\ref{analyticlambda}) (solid
line) for typical
parameters of $m=1/2$, $\Gamma=0.002$, $\omega_0=0.3$, and $g_{\rm ep}=0.10$.
}
\label{figS2}
\end{figure}

We test the approximation leading to Eq.~(\ref{approximate}) by comparing it to
the numerical evaluation of the integral that retains the phase
$qEs|s|/2m$ and the exponential damping factor by $\Gamma$ at $t=0$
\begin{equation}
\lambda'=
\frac{img_{\rm ep}^2}{2\omega_0}
\int\frac{dq}{2\pi}\int ds \frac{e^{-2\Gamma|s|+i(q^2/2m+\omega_0)s
+iqEs|s|/2m}}{qEs+2im\Gamma}
\label{fullint}
\end{equation}
Although the integral converges very slowly due to the extreme
oscillation in the integrand, especially in the low $E$-limit, the
agreement shows that the analytic approximation is reliable in the
regime of interest.

By using the parameter $x=2\Gamma\sqrt{2m\omega_0}/E$ in
Eq.~(\ref{analyticlambda}), we can rewite the
avalanche condition as
\begin{equation}
xK_0(x)=\frac{4}{\beta} \mbox{ with }\beta=\frac{g_{\rm
ep}^2}{\Gamma}\left(\frac{2m}{\omega_0^3}\right)^{1/2},
\end{equation}
with the avalanche parameter $\beta$. In the large $\beta$ limit, the
solution to the criterion can be simplified by using the asymptotic
relation of the Bessel function as
\begin{equation}
\left(\frac{\pi}{2}x\right)^{1/2}e^{-x}\approx\frac{4}{\beta},
\mbox{ and
}x\approx\ln\left(\sqrt{\frac{\pi}{32}}\beta\right)+\frac12\ln x,
\end{equation}
which can be recursively solved as
\begin{equation}
x\approx\ln\left(\sqrt{\frac{\pi}{32}}\beta\right)
+\frac12\ln\left[\ln\left(\sqrt{\frac{\pi}{32}}\beta\right)\right]
+\cdots.
\end{equation}
With the first term, the approximate solution for $E_{\rm av}$ is
\begin{equation}
E_{\rm av}\approx
\frac{2\Gamma\sqrt{2m\omega_0}}{\ln\left(\sqrt{\frac{\pi}{32}}\beta\right)}.
\end{equation}

Here, we give a continued discussion on the energy conservation given in the
main text. As discussed, and also as shown in
Fig.~\ref{figS2}, the integral Eq.~(\ref{analyticlambda}) goes to zero as
$E\to 0$, validating the absence of avalanche mechanism in equilibrium.
Therefore, Eq.~(\ref{analyticlambda}) can be viewed as a deviation from
the energy conservation due to the energy gain of the electron by the
electric field. The integral reaches the maximum at
\begin{equation}
\frac{2\Gamma\sqrt{2m\omega_0}}{E}=x_0=0.595047,
\end{equation}
which can be rewritten as
\begin{equation}
\omega_0=\frac{x_0^2}{4}\cdot\frac{1}{2m}\left(\frac{E}{\Gamma}\right)^2.
\end{equation}
This shows that the integral becomes maximum when the kinetic energy gain
during the time $1/\Gamma$ matches the order of the phonon energy.

\bigskip
\centerline{\bf S2. Derivation of the Kickoff Temperature $T^*$ in $E_{\rm
av}$}
\bigskip

Since the retarded self-energy has the main contribution from the
greater GF, we compute the greater self-energy in the low-field limit as
\begin{equation}
\Sigma^>(t)=-\frac{ig_{\rm ep}^2}{2\omega_0}\int\frac{dp}{2\pi}
e^{-i(\Delta+p^2/2m)t}
[(1+n_b)e^{-i\omega_0t}+n_be^{i\omega_0 t}],
\end{equation}
with the expression inside the bracket due to the phonon propagator.
At the bandedge $\omega=\Delta$, the
term proportional to $(1+n_b)$ vanishes due to the energy conservation,
and we have the Fourier transformation
\begin{equation}
\Sigma^>(\omega=\Delta)\approx
-\frac{ig_{\rm ep}^2 n_b}{2\omega_0}\int dp\,\delta(\omega_0-p^2/2m)
=-\frac{img_{\rm ep}^2}{\omega_0\sqrt{2m\omega_0}}n_b
=-i(\beta/2)\Gamma n_b,
\end{equation}
and ${\rm Im}\Sigma^R(\omega)\approx \frac12{\rm Im}\Sigma^>(\omega)$.
Therefore the effective dephasing rate becomes
\begin{equation}
\Gamma_{\rm eff}=\Gamma\left(1+\frac{\beta}{4}n_b\right).
\end{equation}
We define the activation temperature $T^*$ at the initial rise of $E_{\rm
av}$ and conveniently set for the condition $\beta n_b=1$, which leads to
\begin{equation}
T^*=\frac{\omega_0}{\ln(1+\beta)}.
\end{equation}

\bigskip
\centerline{\bf S3. Lattice Model for Numerical Calculations}
\bigskip

For computational purpose, we discretize the space with the lattice of
a one-band tight-binding chain under a DC
electric-field $E$, in the Coulomb gauge~\cite{liprl2015,ligraphene}
\begin{equation}
H^{\rm 1D}_{0,\rm
el}=\sum_i\left[-t(d^\dagger_{i+1}d_i+d^\dagger_id_{i+1})+(2t+\Delta-Ex_i)d^\dagger_i
d_i\right],
\end{equation}
where $d^\dagger_i/d_i$ is the creation/annihilation operator for an
electron at site $i$, for which the site position $x_i=ia$.  $\Delta$
is the bare gap, $t$ the tight-binding parameter, and $a$ the lattice
constant. To compare with
the continuum model we set $t=1/(2m)=1$. The electrons
are locally coupled to phonons, which are modeled by a collection of
harmonic oscillators given by the Hamiltonian
\begin{equation}
H_{0,\rm ph}=\frac12\sum_i(p_i^2+\omega_0^2\varphi_i^2),
\end{equation}
with $\varphi_i$ the amplitude at site $i$, $p_i$ its momentum,
and $\omega_0$ the phonon frequency. The on-site
electron-phonon coupling is given by
\begin{equation}
H_{\rm ep}=g_{\rm ep}\sum_i\varphi_id^\dagger_i d_i,
\end{equation}
with the coupling constant $g_{\rm ep}$.

We bypass the transient dynamics and directly access the homogeneous nonequilibrium
steady-state of the many-body dynamics~\cite{han_prb2013,liprl2015,ligraphene}. The
fermion baths enter the computation of the electronic Green's function
via local retarded and lesser self-energies at site $i$ as
\begin{equation}
\Sigma^R_{0,i}(\omega)=-i\Gamma,\quad\Sigma^<_{0,i}(\omega)=2i\Gamma
f_0(\omega+Ex_i),
\end{equation}
while the Ohmic baths~\cite{weiss} enter the phonon
Green's function via local self-energies as
\begin{equation}
\Pi^R_{0,i}(\omega)=-2i\tau_P^{-1}\omega,\quad\Pi^<_{0,i}(\omega)=-4i\tau_P^{-1}\omega
n_0(\omega).
\end{equation}
In the above expressions, $f_0(\omega)=(e^{\omega/T}+1)^{-1}$ and
$n_0(\omega)=(e^{\omega/T}-1)^{-1}$ are the Fermi-Dirac and
Bose-Einstein distributions at the bath temperature $T$, respectively,
and $\tau_P$~\cite{khurgin2007} is the phonon decay time. 

We compute the second-order self-energy to electrons
as~\cite{moire2023,ligraphene}
\begin{equation}
\Sigma^\lessgtr_i(\omega) =
ig_{\rm ep}^2\int\frac{{\rm d}\omega'}{2\pi}{\cal
G}^\lessgtr_i(\omega-\omega')D^\lessgtr_\alpha(\omega').
\end{equation}
with ${\cal G}$ the Weiss-field Green's function in the dynamical
mean-field formalism.
The steady-state condition is imposed by setting up the symmetry of the
electron GF~\cite{liprl2015} as
\begin{equation}
G_i(\omega) = G_0(\omega+Ex_i) \mbox{ with }x_i=ia.
\end{equation}



%